\documentclass[12pt,preprint]{aastex}
\usepackage{amsmath}

\begin{document}
\begin{tiny}

\end{tiny}
\title{On the Spin Period Distribution in Be/X-ray Binaries}
\author{Z.-Q. Cheng$^{1}$, Y. Shao$^{1}$, and X.-D. Li$^{1,2}$
}
\affil{$^{1}$Department of Astronomy, Nanjing University, Nanjing 210093, China;
lixd@nju.edu.cn}

\affil{$^{2}$Key laboratory of Modern Astronomy and Astrophysics, Nanjing University,
Ministry of Education, Nanjing 210093, China}


\begin{abstract}

There is a remarkable correlation between the spin periods of the
accreting neutron stars in Be/X-ray binaries (BeXBs) and their
orbital periods . Recently \citet{k2011} showed that the
distribution of the spin periods contains two distinct
subpopulations peaked at $\sim 10$ s and $\sim 200$ s respectively,
and suggested that they may be related to two types of supernovae
for the formation of the neutron stars, i.e., core-collapse and
electron-capture supernovae. Here we propose that the bimodal spin
period distribution is likely to be ascribed to different accretion
modes of the neutron stars in BeXBs. When the neutron star tends to
capture material from the warped, outer part of the Be star disk and
experiences giant outbursts, a radiatively-cooling dominated disk is
formed around the neutron star,  which spins up the neutron star,
and is responsible for the short period subpopulation. In BeXBs that
are dominated by normal outbursts or persistent, the accretion flow
is advection-dominated or quasi-spherical. The spin-up process is
accordingly inefficient, leading to longer periods of the neuron
stars. The potential relation between the subpopulations and the
supernova mechanisms is also discussed.

\end{abstract}

\keywords{binaries: general - stars: neutron - X-rays: binaries}

\section{Introduction}

High-mass X-ray binaries (HMXBs) usually consist of a neutron star
(NS)  and an optical companion star of mass higher than about $8
M_{\odot}$. According to the spectral characteristics of the optical
companions, HMXBs can be further divided into supergiant X-ray
binaries (SGXBs) and Be/X-ray binaries (BeXBs) \citep[][for a recent
review]{Reig2011}. Most BeXBs are transient systems and present
moderately eccentric orbits ($e\gtrsim 0.3$). The NS captures the
wind material from its companion, producing X-ray radiation.
Meanwhile, the spin of the NS changes with time, and both spin-up
and spin-down have been observed when accretion took place
\citep{n1989,b97}.

\citet{Corbet1984,Corbet1985,Corbet1986} first noticed that
different subgroups of HMXBs appear to be located in different
regions in the spin perion ($P_{\rm s})$ vs. orbital period ($P_{\rm
orb}$) diagram (also called the Corbet diagram). In particular,
there exists a positive correlation between $P_{\rm s}$ and $P_{\rm
orb}$ for BeXBs, although with a large observed scatter. The
relations between $P_{\rm s}$ and $P_{\rm orb}$ in HMXBs are likely
to reflect the wind structure and accretion processes in HMXBs
\citep[e.g.,][]{swr1986,hr87,wv1989,king1991,lv1996}.

It is generally thought that the interaction between the NS magnetic
field and the captured material from its binary companion can lead
to a so-called equilibrium spin period $P_{\rm eq}$ of the NS
\citep{bh91}. However, the derived values of $P_{\rm eq}$ for
wind-accreting SGXBs are always lower than the observed ones of
$P_{\rm s}$. It was suggested that the present $P_{\rm s}$
distribution may result from the equilibrium spin period when the
companion star was still on the main sequence with a much weaker
wind, and the wind of a SG is unable to transfer enough angular
momentum to move the NS towards a new equilibrium value (e.g. Stella
et al. 1986). The situation is more complicated in BeXBs. Be star
winds are known to be disklike rather spherically expanding as in
SGs, and a Be star can transform to be a B star and vice versa from
time to time. The mechanism for this transition is still unknown.
The varying Be star wind and the eccentric orbit imply that there
does not exist  a stable equilibrium spin period. \citet{wv1989}
argued that the observational selection effects, that is, BeXBs are
more likely to be observed when the NS moves within the dense
equatorial disk wind, and the difference between the spin-up and
spin-down timescales when the NS accretes within and outside of the
disk wind, imply that $P_{\rm s}$ is potentially correlated with the
accretion rate during outbursts, and thus $P_{\rm orb}$.

Recently \citet{k2011} showed that the $P_{\rm s}-P_{\rm orb}$
correlation in BeXBs becomes
more dispersed and a bimodal distribution for both $P_{\rm s}$ and $P_{\rm orb}$
seems to exist. While the bimodality is somewhat marginal in $P_{\rm orb}$,
the $P_{\rm s}$ distribution has a clear gap at $\sim 40$ s with two
peaks around 10 s and 200 s, respectively. \citet{k2011} proposed that two types of supernovae
(SNe) may be responsible for the two subpopulation of BeXBs:
the electron-capture supernovae (ECS)
usually produce NSs with shorter spin periods, and
lower eccentricities, while iron core-collapse
supernovae (CCS) are preferred for NSs
with longer spin periods  and higher eccentricities.

The original idea for the ECS candidates in BeXBs stems from a
subclass of BeXBs that can be explained by low NS kicks
\citep{p2002}. These BeXBs are characterized by long spin periods,
persistent low X-ray luminosities ($\sim 10^{34}-10^{35}$
ergs$^{-1}$),  wide binary orbits ($P_{\rm orb}> 30$ days), and low
eccentricities ($e\lesssim 0.2$). The prototype of them is X Persei
\citep[$P_{\rm s}=837$ s,][]{w1976}. Other sources include RX
J0146.9$+$6121 \citep[1412 s,][]{h1998}, RX J1037.5$+$5647 (860 s)
and RX J0440.9+4431 (202.5 s) \citep{rr1999}. Recently discovered
BeXBs SXP 1062 \citep[1062 s,][]{h2012}, 1RXS J225352.8$+$624354
\citep[47 s,][]{e2013}, and SWJ2000.6$+$3210 \citep[890 s,][]{p2013}
may also belong to this subclass. \citet{p2004} and \citet{vdh2004}
suggested that the ECS mechanism may account for the low kicks in
these BeXBs. This is different from the proposal by \citet{k2011}
that ECS-BeXBs may have short spin periods, relatively narrow orbits
and low eccentricities.

A thorough investigation on the origin of the subpopulations of
BeXBs requires a population synthesis incorporating stellar and
binary evolution, SN explosions, the Be star wind structure, mass
transfer processes, and the NS evolution, which is beyond the scope
of this paper. Here we focus on the origin of the the $P_{\rm s}$
distribution, which shows a much clearer bimodal feature than the
$P_{\rm orb}$ distribution in the current sample. We expect that the
orbital periods of BeXBs may be largely dependent on the initial
parameters of the progenitor binaries and the SN mechanisms, since
tidal interaction is unable to change them effectively in wide
orbits, while the NS spin periods are likely to be determined by the
accretion processes in BeXBs.
In Section 2 we compare the statistical characteristics of the
outbursts in the two subpopulations, and present qualitative
argument that the bimodal $P_{\rm s}$ distribution can be ascribed
to different accretion modes of the NSs. We discuss the possible
implications on the SN mechanisms in Section 3, and summarize in
Section 4.

\section{The spin period distribution}

It is uncertain how different types of SNe can influence the initial parameters
of the newborn NSs. However, since the current spin periods of the NSs in BeXBs are
generally much longer than
the initial periods, their distribution must be determined
by the interaction between the NSs and the captured material during the
evolution, which has erased any feature in the initial distribution.
So the bimodal $P_{\rm s}$ distribution is likely to result from different
spin histories of the NSs.

The NS spin evolution in HMXBs depends on the angular momentum transfer
between the NS and the captured matter from its companion star.
It can be briefly outlined as follows \citep[][see also Dai et al. 2006]{dp1981}.
A newborn NS usually spins rapidly (with $P_{\rm s}$ much less than 1 second)
so that the transferred matter from
its companion star is stopped by the strong magnetic dipole radiation
outside the light cylinder radius (or the Bondi accretion radius).
The NS acts as a radio pulsar
with magnetic dipole radiation responsible for its spin-down.
This is called the ejector phase.
When the magnetic dipole radiation can no longer prevent the wind matter
penetrating the light cylinder, the accretion flow starts to interact with the NS
magnetic field, ceasing the pulsar activity, and the NS enters the propeller phase.
In this phase, the accretion flow is balanced by the rotating
magnetic field at the magnetospheric radius $R_{\rm m}$.
Accretion is inhibited and the rotating NS loses its angular momentum by ejecting the material
at $R_{\rm m}$.
This propeller phase ends when the accretion flow overcomes the centrifugal barrier and
falls onto the NS. At this time the NS spin period evolves to the equilibrium period
given by \citep{dp1981}
\begin{equation}
P_{\rm eq,w}\simeq 62B_{12}^{6/7}R_6^{18/7}
M_1^{-5/7}\dot{M}_{14}^{-3/7}\,{\rm s},
\end{equation}
where $B=10^{12}B_{12}$ G is the NS surface magnetic field strength,
$M=M_1M_{\sun}$ the NS mass, $R=10^6R_6$ cm the NS radius, and
$\dot{M}=10^{14}\dot{M}_{14}$ gs$^{-1}$ the mass capture rate.
If the value of $B$ is constant, the maximum of $P_{\rm eq,w}$ is attained
when $\dot{M}$ takes its lowest value. This occurs when the Be star wind changes
from disklike to spherical and/or  the NS
moves around the apastron in an eccentric orbit.
In the following accretor phase, if the captured material possesses
enough angular momentum it will evolve into
an accretion disk, so the NS can be spun up or down
to a new equilibrium period \citep{pr1972,gl1979},
\begin{equation}
P_{\rm eq,d}\simeq 4.8(\omega_{\rm c}/0.5)^{-1}B_{12}^{6/7}R_6^{18/7}
M_1^{-5/7}\dot{M}_{17}^{-3/7}\,{\rm s},
\end{equation}
where $\omega_{\rm c}$ is the ``fastness" parameter ranging between 0 and 1,
and $\dot{M}_{17}=\dot{M}/10^{17}$ gs$^{-1}$.
Since most BeXBs are in eccentric orbits and transient, the NS
does not evolve along the above track monotonously, but transits between the
propeller (sometimes even the ejector) and the accretor phases, with its spin period lying
between $P_{\rm eq,w}$ and $P_{\rm eq,d}$. If there is efficient disk accretion
and spin-up, $P_{\rm s}$ is likely to be close to $P_{\rm eq,d}$. Otherwise, it
stays around $P_{\rm eq,w}$.

Observational and theoretical developments since 1990s have shown that
the winds from Be stars are in the form of Keplerian disks, held by viscosity
and having small radial velocities
\citep[e.g.,][]{l1991,w1997,o2001,pr2003,c2009,j2009,s2009,m2013}.
Because of its eccentric orbit, the NS can capture gas from the Be star
disk only for a short span of time when it moves close to the disk,
giving rise to transient X-ray outbursts.
There are two types of X-ray outbursts in BeXBs.
Normal (Type I) X-ray outbursts  occur at or near the periastron passage.
The X-ray luminosity ($L_{\rm X}$) increases from its quiescent value
by about one order of magnitude
to $\sim 10^{36}-10^{37}$
ergs$^{-1}$. The duration is a small fraction of the
orbital period, typically $(0.2 - 0.3) P_{\rm orb}$.
Giant (Type II) outbursts are significantly
brighter ($L_{\rm X} >10^{37}$ ergs$^{-1}$) and less frequent than normal
outbursts. The duration is about tens of days ($\gtrsim 0. 5P_{\rm orb}$, sometimes
over one orbital period), and no orbital modulation has been detected.
Spin-up episodes of the NS in BeXBs have been seen during both
giant and normal outbursts \citep[e.g.,][]{p1989,b97,w2008}, suggesting
the existence of an accretion disk.

The detailed process of how a NS captures matter from
the Be star disk is not clear. Angular momentum transfer in this regime is also
difficult to quantify, though some information can be obtained from the SPH simulations
\citep{ho2004,ho2006}.
\citet{no2001} and \citet{on2001} argued that, the Be star disk is tidally truncated
by the NS so that mass transfer occurs preferentially via
leakage from the disk at the inner Lagrangian point near the periastron passage \citep{o2002},
resulting in normal outbursts. Giant outbursts are likely to be
caused by accretion  from a warped Be star disk
that is misaligned with the binary orbital plane \citep{n2001,o2013}.
The reason is that, in misaligned systems the
tidal torque is weaker than in coplanar systems, so the truncation radius
could be larger than the periastron separation, and the NS
could capture material at a high enough rate
when passaging through the warped part of the Be disk \citep{m2011}.
There is observational evidence for warped Be
star disks before or during giant outbursts. For example,
\citet{n2001} and \citet{r2007} found that before and during the giant outbursts
of 4U 0115$+$634, the H$\alpha$ emission line
from the optical counterpart  changed from a usual double-peaked
profile to a single-peaked or shell-line profile on a timescale of a year or so.
Complicated changes in the H$\alpha$ line profiles during and after the 2009
giant outburst of A0535$+$262 have also been observed \citep{mo2011}, and interpreted
by a precessing, warped Be star disk  \citep{mo2013}.
The misalignment between the spin axes of the Be star and
the binary orbit is thought to originate from
the SN explosion, especially the associated kick \citep{m2009}.

To investigate whether there is any potential relation between the outburst characteristics
and the bimodal $P_{\rm s}$ distribution, we plot in Fig.~1 the distribution of
BeXBs in the Galaxy, the Large and Small Magellanic
Clouds (LMC and SMC) with known outburst behavior.
The green, red, blue and orange symbols denote
sources with type I outbursts only, with type II outbursts only,
persistent sources, and sources with both type I and type II outbursts,
respectively \citep[data are taken from][and references therein]{rp05,r2011,t2011}.
{The dashed horizontal and vertical lines correspond to $P_{\rm s}=40$ s
and $P_{\rm orb}=60$ days, which separate the population in $P_{\rm s}$
and $P_{\rm orb}$, respectively.
Among the 30 short-$P_{\rm s}$ BeXBs, 23 showed giant outbursts,
while only 10 out of the 39 long-$P_{\rm s}$ BeXBs showed giant outbursts.
This seems to indicate that the $P_{\rm s}$ distribution may be related
to the occurrence of giant outbursts, which are characterized
by long episodes of spin-up.
We also plot the same distribution in Fig.~2,
discriminating systems with different peak luminosity (in units of
$10^{37}$ ergs$^{-1}$) during outbursts. Obviously short-$P_{\rm s}$ systems 
tend to have higher peak luminosities.
To show this feature more clearly, we plot in
Fig.~3 the distribution of the peak (or persistent) luminosity for BeXBs
with $P_{\rm s}>40$ s (blue line) and $\le 40$ s (red line).
A bimodal distribution with characteristic luminosities of a few $10^{36}$
ergs$^{-1}$ and a few $10^{38}$ ergs$^{-1}$ is seen.
Note that this result may be subject to the difference in
the star formation histories and the metallicities of the Galaxy, 
LMC and SMC, which seem to play a role in the formation
of Be stars, as they are more frequent
in the SMC.
Although a complete census of the outbursts in BeXBs is lacking, these
figures indicate that the current spin period distributions in the two
subpopulations are likely related to the different outburst characteristics.

However, the difference in the NS accretion rate itself is not
large enough to explain the difference in the spin periods in the two subpopulations.
According to Eq.~(1) or (2), changing $\dot{M}$ by a factor of $10-100$ can lead to
the change of $P_{\rm s}$ by a factor of $\sim 3-7$.
A more important factor may lie in different accretion modes of the
NSs. Although during both normal and giant outbursts there may
be (transitional) accretion disks formed around the NSs, the structure
of the disks can be quite different.
It was pointed out by \citet{o2013} that, if the accretion disk
during normal outbursts is a geometrically thin, radiatively-cooling dominated \citep{ss1973},
the accretion timescale (i.e., the viscous timescale) will be several times
the orbital period, much longer than the outburst duration.
In such a situation the system will exhibit no rapid nor
large X-ray flux changes seen in the outbursts.
This implies that the accretion flow should be radiatively inefficient.
Indeed, it is already known that accretion disks with
$\dot{M}<\sim 10^{16}$ gs$^{-1}$  are likely to be in the form of
geometrically thick, advection-dominated accretion flows (ADAFs) \citep{ny1994,ny1995}.
The radial velocity in ADAFs is comparable with the Keplerian
velocity, so that the accretion timescale is much shorter than in
thin disks, consistent with the duration of normal outbursts.
Meanwhile the angular velocity in ADAFs is significantly
lower than the Keplerian one in thin disks.
The consequence is that the spin-up torque is relatively smaller and
the equilibrium period becomes longer \citep{dl2006}, i.e.,
\begin{equation}
P_{\rm eq,ADAF}\simeq 64.4 (A/0.2)^{-1}(\omega_{\rm c}/0.5)^{-1}B_{12}^{6/7}R_6^{18/7}
M_1^{-5/7}\dot{M}_{16}^{-3/7}\,{\rm s},
\end{equation}
where $A=\Omega_{\rm ADAF}(R_{\rm m})/\Omega_{\rm K}(R_{\rm m})$, and
$\dot{M}_{16}=\dot{M}/10^{16}$ gs$^{-1}$. Taking typical value of $A\sim
0.2-0.3$ \citep{y1997}, we find that the equilibrium period $P_{\rm eq,ADAF}$
is several times larger than $P_{\rm eq,d}$ with the same values of $B$ and $\dot{M}$.

According to the above arguments, we tentatively propose an explanation for
the bimodality in the $P_{\rm s}$ distribution. In BeXBs that tend to
experience giant outbursts, the NS accretes from a thin disk with relatively
long lifetime,
the efficient mass and angular momentum transfer results in spin-up during
the outbursts, so that its spin period reaches $\sim P_{\rm eq,d}$ with
typical value of $\sim 10$ s. In BeXBs where
normal outbursts dominate or  there are no outbursts at all, the accretion flow
around the NS is an ADAF (or quasi-spherical), so that
there is relatively infrequent effective spin-up. The spin period lies between
$\sim P_{\rm eq,w}$ and $\sim P_{\rm eq,ADAF}$, with typical value
of $\sim 100$ s. Thus the subpopulations
of BeXBs may originate from different accretion modes of the NSs.

In the above picture, disk warping plays an important role in determining
the spin evolution of the NSs. In the Be star disks the tidal torque
exerted by the NS balances the viscous
torque at the tidal warp radius $R_{\rm tw}$. Outside of $R_{\rm tw}$ the disk
is dominated by the tidal torque,  which flattens the disk and makes it to align with
the binary orbital plane \citep{m2009}\footnote{If the disk does not extend up to the tidal
radius, the torque can still have an effect on the disk and cause the
it to move towards alignment with the binary orbit even if they
do not completely align.}. With standard parameters (i.e., $1.4\,M_{\sun}$ NS and
$17\,M_{\sun}$ Be star) for a BeXB, the tidal warp radius
can be expressed as follows \citep{m2011},
\begin{equation}
R_{\rm tw}\simeq 9.3\times 10^{10}(1-e^2)^{3/4}\alpha
(\frac{P_{\rm orb}}{1\,{\rm d}}) \,{\rm cm},
\end{equation}
where $\alpha$ is the viscous parameter corresponding to the vertical shear
in the disk. For BeXBs, the mass ratio of the NS and the Be star
is approximately $0.1-0.2$, so the truncation
radius is $\lesssim 0.5d$, where $d$ is the binary separation.
 A necessary condition for the interaction between a warped disk
 and a NS at periastron is $R_{\rm tw}<0.5a(1-e)$,
where $a$ is the semi-major axis of the binary. 
Combining this with Eq.~(4) yields
\begin{equation}
P_{\rm orb}<(72.5\,{\rm days}) \alpha^3\frac{(1-e)^3}{(1-e^2)^{9/4}}.
\end{equation}
In Fig.~4 we show the distribution of BeXBs in the $P_{\rm orb}-e$
diagram, with blue, red, and green symbols representing
binaries with $P_{\rm s}>40$ s, $\le 40$ s, and unknown spin periods, respectively.
The two curves correspond to the limit given by Eq.~(5) with $\alpha=1$ (upper)
and 0.5 (lower). We see that most of the short-$P_{\rm s}$ BeXBs are in
the regions confined by the curves, suggesting possible existence
of disk warping.

\section{Discussion}

We emphasize that the argument in Section 2 is based on the
global properties of
the BeXB population rather than the individual source characteristics.
Actually Fig.~1 shows that there is no strict one to one correspondence between
short/long $P_{\rm s}$ and the occurrence
of giant/normal outbursts. The transient nature of
BeXBs means that it is impossible to monitor all outbursts
for each source, so our classification of the outburst
behavior is obviously incomplete. Evolution of the Be star disk also influences
the characteristics of outbursts and the NS spin evolution.
Nevertheless, the features in Figs.~1- 3 strongly suggest that different
accretion modes of the NSs may be behind the origin of the subpopulations.

The SN mechanisms can be related with the subpopulations through their
influence on the initial orbital period, eccentricity, and misalignment
of the Be star disk.
Population synthesis calculations by \citet{l2009} showed that the
ECS channel may be efficient at forming BeXBs, especially in the SMC, in which
the population of HMXBs has been found to have relatively wide orbits
and low eccentricities.
However, they did not compare the characteristics of BeXBs formed
through the CCS and ECS channels.
To examine the influence of different kinds of SNe on the orbital period distribution,
we employ a Monte-Carlo method to simulate the formation of
BeXBs. We adopt the binary population synthesis (BPS)
code developed by \citet{h2000,h2002} to
calculate the evolutions of a large number of the primordial binaries, which is
similar to the code {\em StarTrack} used by \citet{l2009}.
We consider Solar abundance for the stars, with
most of the input parameters (i.e., the distributions of the orbital separation, mass ratio,
the initial mass function  of the mass of the primary star)  same as the standard ones
described by \citet{h2002}. Detailed description of the method and calculated results will
be presented elsewhere \citep{sl2014}. Some relevant key points in the calculations
are listed below.


\begin{enumerate}
\item We consider both CCS and ECS for the NS formation.
For ECS, we adopt the following criterion suggested by \citet{f12}.
If the core mass $ M_{\rm c,bagb}$ of the primary star (i.e., the NS's progenitor)
at the base of asymptotic giant branch is between
$ 1.83 M_{\odot} $ and $ 2.25 M_{\odot} $, the CO core will non-explosively
burn into an ONe core, and the core mass is  accumulated gradually.
If its mass can reach $ M_{\rm esc} = 1.38 M_{\odot} $, the ONe core
will collapse due to electron capture into Mg and form a NS. If the
mass is less than $  M_{\rm esc} $, it will leave an ONe WD.

\item We apply a Maxwellian distribution for the SN kick velocity
imparted to the newborn NS, with one dimensional rms velocity $\sigma=265$ kms$^{-1}$ for CCS
\citep{h2005} and $50$ kms$^{-1}$ for ECS \citep{p2002}, respectively.

\item We define a BeXB to be a binary consisting of  a NS and
a main-sequence (i.e., core H burning) companion star
with mass between 8 $M_{\odot}$ and 20 $M_{\odot}$, which does not fill its
Roche-lobe. Since Be stars
are rapidly rotating, we also consider the influence of tidal synchronization on the Be stars.
Only systems with the synchronization timescale greater than the main-sequence
lifetime of the Be star are taken into account.
\end{enumerate}

The calculated normalized orbital period distributions of BeXBs are plotted in Fig.~5.
The red and black curves denote systems formed through CCS and ECS, respectively.
In the left panel the secondary star is regarded as a Be star
when its rotational velocity is accelerated to
$80\%$ of its break-up velocity due to previous mass transfer. In
the right panel we assume that a constant
fraction of B stars are Be stars. Note that in both cases the orbital periods
have similar, wide distributions, with CCS-BeXBs peaked at $P_{\rm orb}\sim 40-50$
days and ECS-BeXBs at  $\sim 100$ days. It seems that CCS-BeXBs dominate
at $P_{\rm orb}<\sim 20-50$ days, but at longer $P_{\rm orb}$ the numbers of
the two classes of objects are comparable.
We need to caution that
the number and distribution of ECS-BeXBs
depend on the adopted mass range of the ECS progenitors
\citep{n1984,n1987,p2004,s2007,p2008}, which is not well understood.
Nevertheless, a bimodal orbital period distribution from the two types of SN
channels seems not to exist.

Another  factor that can influence the $P_{\rm s}$ distribution and
might be related to the SN mechanisms is the NS magnetic field.
In the above estimates we assume that all the NSs possess a
magnetic field of order $10^{12}$ G. It is not known how NSs born
in CCS and ECS differ in their magnetic fields. For BeXBs with
measured cyclotron resonance scattering features in their X-ray spectra,
the characteristic line energies range from 10 to 55 keV \citep[][and references therein]{p2012},
suggesting comparable field strengths (a few $10^{12}$ G)
in the NSs. However, there is indirect
evidence that the NS magnetic field strengths may occupy a wide range.
For example, from the measured spin-up rate in the 9.28 s Be/X-ray pulsar 2S1553$-$542,
\citet{pp2012} derived a relatively low field $B\sim 5\times 10^{11}$ G
for the NS. In another case, the spin-down rate measured in
the 1062 s  Be/X-ray pulsar SXP1062 implies that the NS possesses a magnetic field
$B\gtrsim 10^{14}$ G \citep{fl2012}. The large scatter of BeXBs in the Corbet
diagram may be partly due to the distribution and evolution of the NS magnetic field.

\section{Summary}
In this paper we argue that the bimodal $P_{\rm s}$ distribution in BeXBs may not
be directly linked to the two SN channels  for the NS formation,
and is more likely to be ascribed to
the difference in the accretion flows onto the NSs. This is indicated by
the occurrence of giant/normal outbursts in the short- and long-$P_{\rm s}$ subpopulations,
which reflect different spin-up efficiencies.
We point out that the difference in the accretion rate
during normal and giant outbursts is not enough to account for
the range of $P_{\rm s}$ in the two subpopulations, and the structure of the accretion
flows may play a more vital role.
Normal outbursts are thought to be triggered by the mass transfer from a tidally
truncated disk at or near periastron passage while giant outbursts are
somehow associated with the warping episodes of the Be star disk
\citep[][and references therein]{o2013}.
On one hand, if the accretion disk during the normal outbursts are transitional and
in the form of ADAF, the NS spin period is relatively longer
due to the lower accretion rate, shorter spin-up duration, and especially lower angular velocity
in the ADAF (the persistent low-luminosity BeXBs usually have long
spin periods, and the NSs may be fed by low-velocity winds).
On the other hand, the accretion disks formed
during giant outbursts are radiatively-cooling dominated, and
the NSs experience longer episodes of spin-up with higher accretion rate,
evolving to shorter spin periods. The two types of SN mechanisms can influence the
NS spin evolution through the configuration of the Be star disk, but
they seem not to result in a bimodal distribution of the orbital period.

\begin{acknowledgements}
We are grateful to an anonymous referee for helpful comments.
This work was supported by the Natural Science Foundation of China
under grant numbers 11133001 and 11333004.

\end{acknowledgements}

\newpage

\begin{figure}[h,t]
\centerline{\includegraphics[angle=0,width=1.00\textwidth]{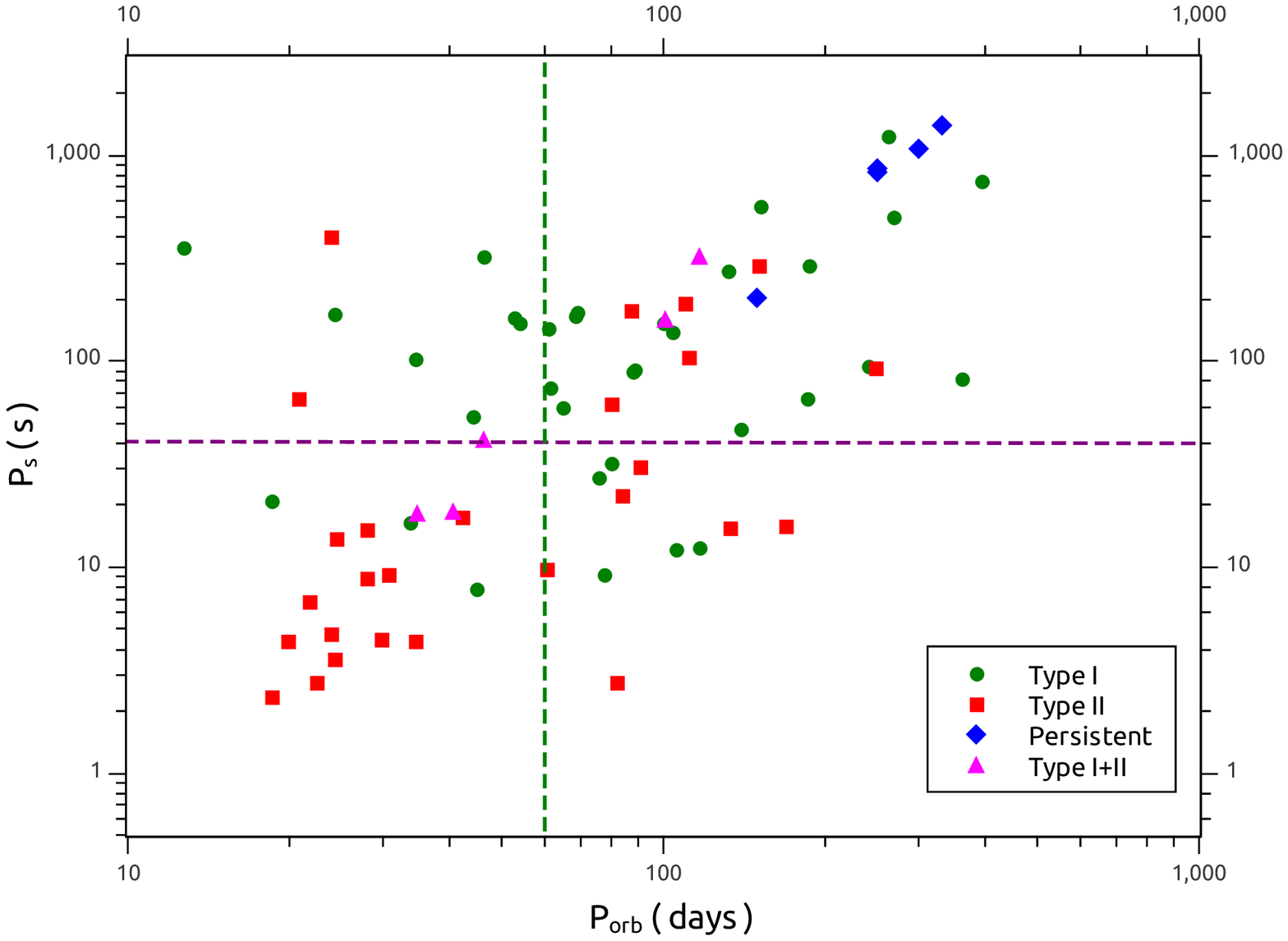}}
\caption{The Corbet diagram for BeXBs. We use different colored symbols to
represent the outburst behavior for each source. The dashed horizontal
and vertical lines correspond to $P_{\rm s}=40$ s
and $P_{\rm orb}=60$ days, respectively.
\label{figure3}}
\end{figure}

\begin{figure}[h,t]
\centerline{\includegraphics[angle=0,width=1.00\textwidth]{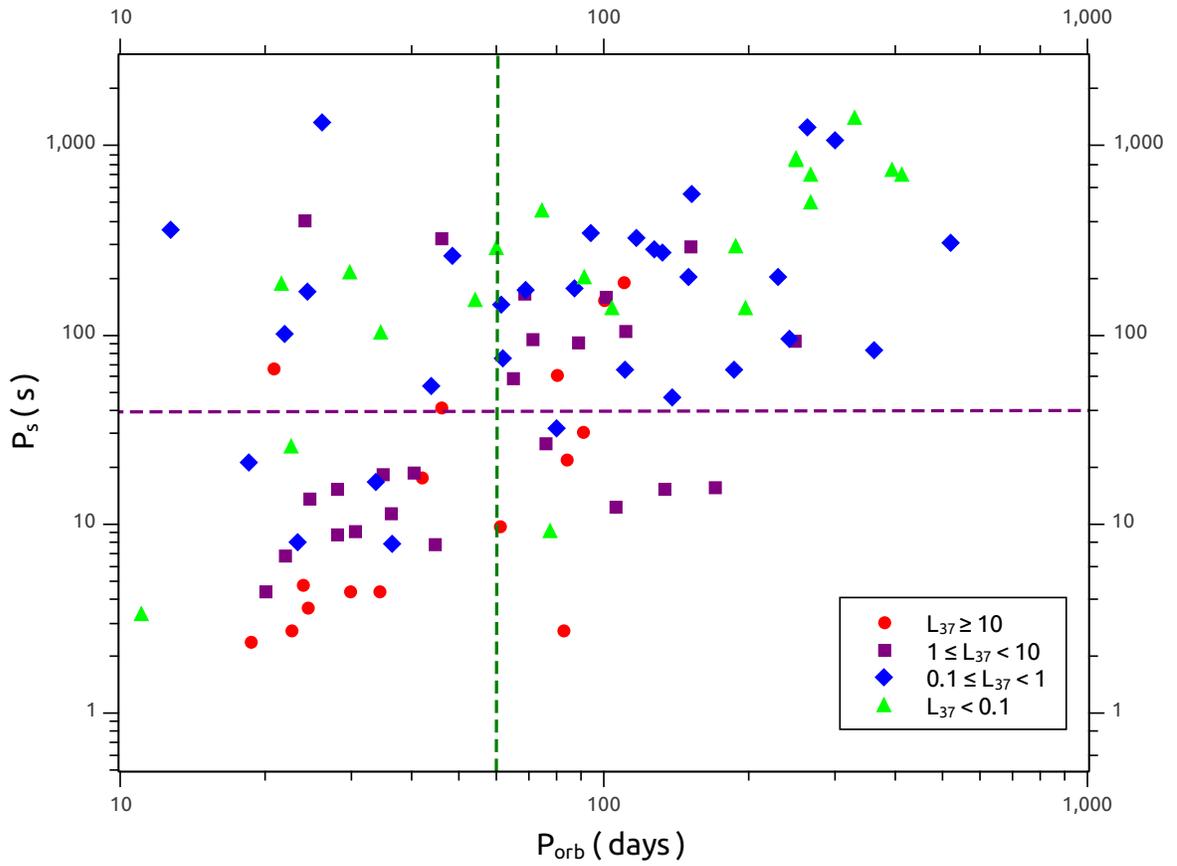}}
\caption{The Corbet diagram for BeXBs. We use different colored symbols to
represent the range of the peak luminosity for each source.
\label{figure4}}
\end{figure}

\begin{figure}[h,t]
\centerline{\includegraphics[angle=0,width=1.00\textwidth]{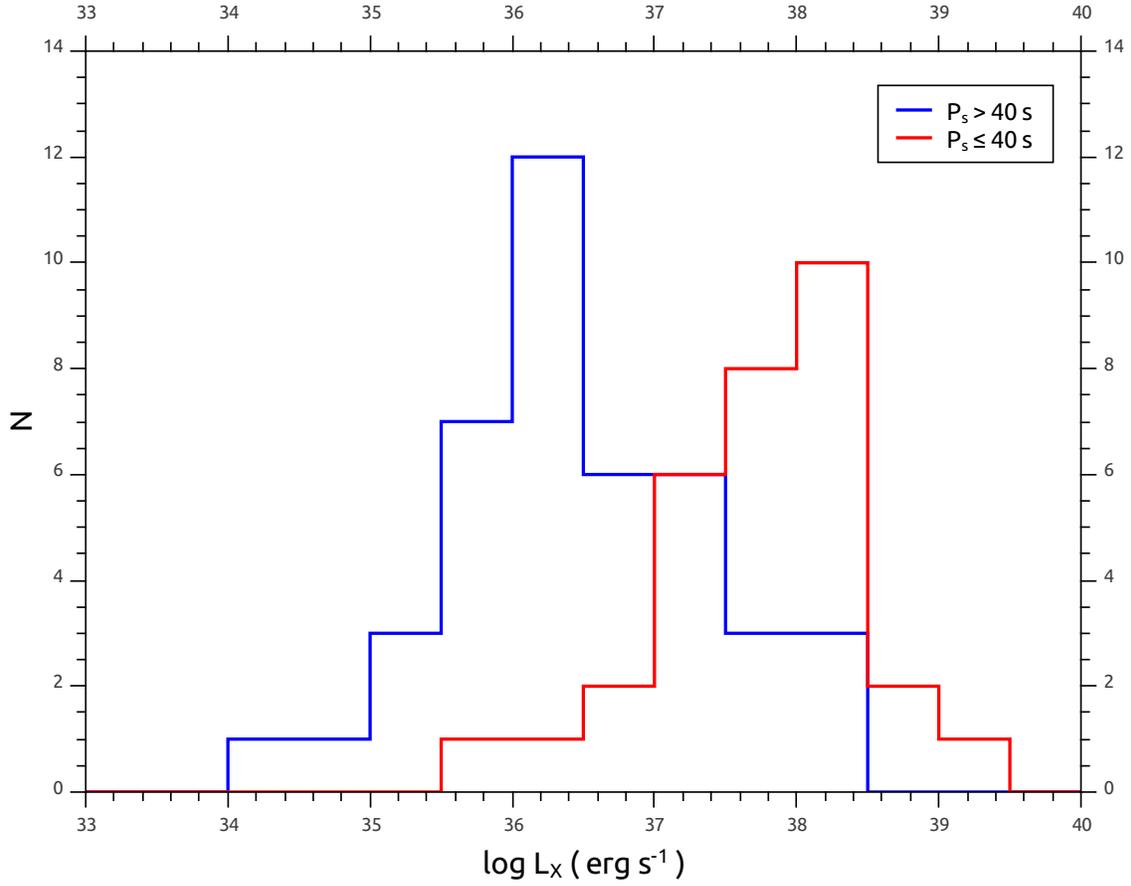}}
\caption{The histogram distribution of the peak (or persistent) luminosities
of BeXBs. The
blue and red lines denote systems with the NS spin periods
$> 40$ s and $\leq 40$ s, respectively.
\label{figure5}}
\end{figure}

\begin{figure}[h,t]
\centerline{\includegraphics[angle=0,width=1.00\textwidth]{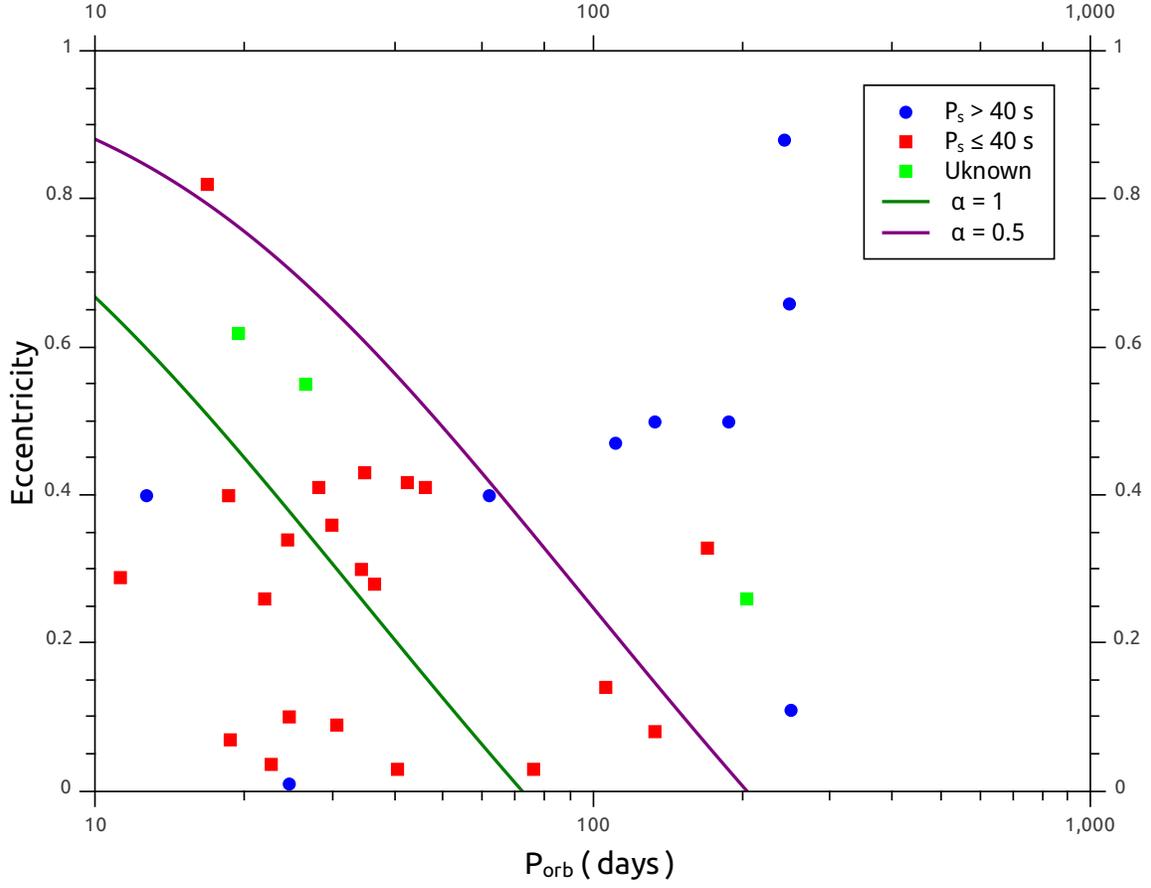}}
\caption{The dependence of the eccentricity on the orbital period in BeXBs.
The blue, red, and green symbols denote systems with the NS spin periods
$> 40$ s, $\leq 40$ s, and with unknown spin periods, respectively.
The two curves correspond to Eq.~(5) with $\alpha=1$ (upper)
and 0.5 (lower), respectively.
\label{figure2}}
\end{figure}

\begin{figure}[h,t]
\centerline{\includegraphics[angle=0,width=1.00\textwidth]{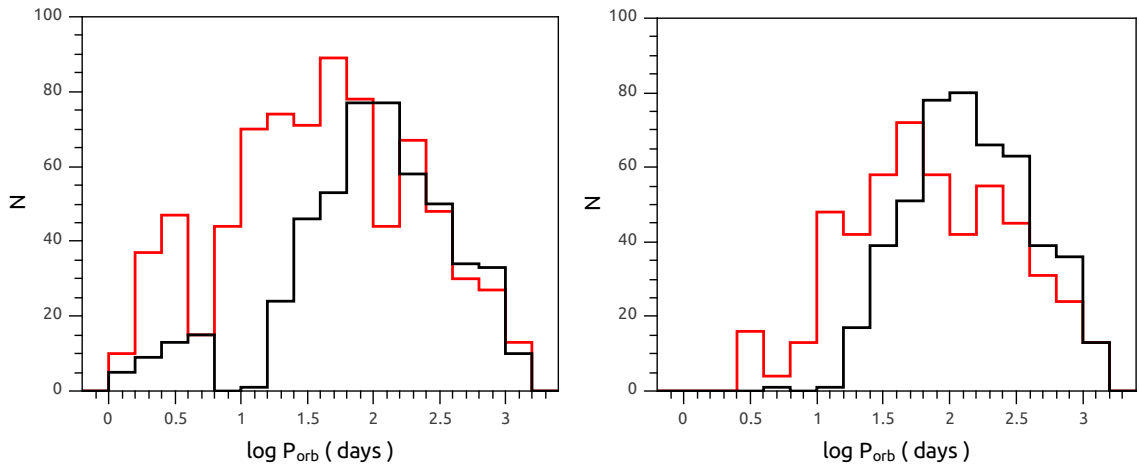}}
\caption{The normalized orbital period distribution of BeXBs derived
from BPS calculations. The red and black lines denote systems in which
the NSs were born with CCS and ECS, respectively. In the left and right
panels, the Be stars are assumed to be rapidly rotating B stars caused
by previous mass transfer, and occupy a constant fraction of B stars,
respectively.
\label{figure1}}
\end{figure}

\clearpage

\label{lastpage}

\begin{thebibliography}{28}

\expandafter\ifx\csname natexlab\endcsname\relax\def\natexlab#1{#1}\fi

%

\bibitem[{{Bhattacharya \& van den Heuvel}(1991)}]{bh91}
Bhattacharya, D. \& van den Heuvel, E. P. J. 1991, Phys. Rep., 203, 1

\bibitem[{{Bildsten et al.}(1997)}]{b97}
Bildsten, L., Chakrabarty, D., Chiu, J., et al. 1997, \apjs, 113, 367

%
%

\bibitem[{{Carciofi et al.}(2009)}]{c2009}
Carciofi, A. C.,  Okazaki, A. T., Le Bouquin, J.-B., et al. 2009, \aap, 504, 915

\bibitem[{{Coe et al.} (2010)}]{c2010}
Coe, M.J., McBride, V. A., \& Corbet, R. H. D. 2010, \mnras, 401, 252

\bibitem[{{Corbet}(1984)}]{Corbet1984}
Corbet, R. H. D. 1984, \aap, 141, 91

\bibitem[{{Corbet}(1985)}]{Corbet1985}
Corbet, R. H. D. 1985, \ssr, 40,409

\bibitem[{{Corbet}(1986)}]{Corbet1986}
Corbet, R. H. D. 1986, \mnras, 220, 1047

\bibitem[{{Dai \& Li}(2006)}] {dl2006}
Dai, H.-L. \& Li, X.-D. 2006, \aap, 451, 581

\bibitem[{{Dai et al.}(2006)}] {dll2006}
Dai, H.-L., Liu, X.-W.,  \& Li, X.-D. 2006, \apj, 653, 1410

\bibitem[{{Davies \& Pringle}(1981)}] {dp1981}
Davies, R. E. \& Pringle, J.E., 1981, \mnras, 196, 209

%
%
%
%
%
%
%
\bibitem[{{Esposito et al.}(2013)}]{e2013}
Esposito, P.,  Israel, G. L., Sidoli, L., et al. 2013, \mnras, 433, 2028

\bibitem[{{Foglizzo, Galletti, \& Ruffert}(2005)}]{fgr2005}
Foglizzo, T., Galletti, P., \& Ruffert, M. 2005, \aap, 435, 397

\bibitem[{{Fryer et~al.}(2012)}]{f12}
Fryer, C., Belczynski, K., Wiktorowicz, G., et al. 2012, \apj, 749, 91

\bibitem[{{Fu \& Li}(2012)}]{fl2012}
Fu, L. \& Li, X.-D. 2012, \apj, 757, 171

\bibitem[{{Ghosh \& Lamb}(1979)}] {gl1979}
Ghosh, P. \& Lamb, F. K., 1979, \apj, 234, 296

\bibitem[{{Haberl et al.}(1998)}] {h1998}
Haberl, F., Angelini, L., \& Motch, C. 1998, \aap, 335, 587

\bibitem[{{Haigh et al.}(2004)}] {h2004}
Haigh, N. J., Coe, M. J., \& Fabregat, J. 2004, \mnras, 350, 1457

\bibitem[{{Haigh \& Okazaki}(2004)}] {ho2004}
Hayasaki, K. \& Okazaki, A. T. 2004, \mnras, 350, 971

\bibitem[{{Haigh \& Okazaki}(2006)}] {ho2006}
Hayasaki, K. \& Okazaki, A. T. 2006, \mnras, 372, 1140

\bibitem[{{H\'enault-Brunet et al.}(2012)}] {h2012}
H\'enault-Brunet, V., Oskinova, L. M., Guerrero, M. A., et al. 2012, \mnras, 420, L13

\bibitem[{{Hobbs et al.}(2005)}] {h2005}
Hobbs, G., Lorimer, D. R., Lyne, A. G., \& Kramer, M. 2005, \mnras, 360, 974

\bibitem[{{Hurley et al.}(2000)}]{h2000}
Hurley, J. R., Pols, O.R., \& Tout, C. A. 2000, \mnras, 315, 543

\bibitem[{{Hurley et al.}(2002)}]{h2002}
Hurley, J. R., Tout, C. A., \& Pols, O. R. 2002, \mnras, 329, 897

%
%
%
%
\bibitem[{{Jones et al.}(2009)}] {j2009}
Jones, C. E., Molak, A., Sigut, T. A. A., et al. 2009, \mnras, 392, 383
%
\bibitem[{{King}(1991)}] {king1991}
King, A. R. 1991, \mnras, 250, 3

\bibitem[{{Knigge et al.}(2011)}] {k2011}
Knigge, C., Coe, M. J., \& Podsiadlowski, Ph. 2011, \nat, 479, 372

%

\bibitem[{{Lee et al.}(1991)}]{l1991}
Lee, U., Osaki, Y., \& Saio, H. 1991, \mnras, 250, 432

\bibitem[{{Li \& van den Heuvel}(1996)}] {lv1996}
Li, X.-D. \& van den Heuvel, E. P. J. 1996, \aap, 314, L13

\bibitem[{{Linden et al.}(2009)}]{l2009}
Linden, T., Sepinsky, J. F., Kalogera, V., \& Belczynski, K. 2009, \apj, 699, 1573


\bibitem[{{Martin et al.}(2011)}] {m2011}
Martin, R. G., Pringle, J.E., Tout, C. A., \& Lubow, S. H. 2011,
\mnras, 416, 2827

\bibitem[{{Martin et al.}(2009)}] {m2009}
Martin, R. G., Tout, C. A., \& Pringle, J. E. 2009, \mnras, 397, 1563

\bibitem[{{McGill et al.}(2013)}] {m2013}
McGill, M. A., Sigut, T. A. A., \& Jones, C. E. 2013, \apjs, 204, 2

\bibitem[{{McSwain \& Gies}(2005)}]{mg2005}
McSwain, M. V. \& Gies, D. R. 2005, \apjs, 161, 118

\bibitem[{{Moritani et al.}(2011)}] {mo2011}
Moritani, Y., Nogami, D., Okazaki, A. T.,  et al. 2011, \pasj, 63, L25

\bibitem[{{Moritani et al.}(2013)}] {mo2013}
Moritani, Y., Nogami, D., Okazaki, A. T.,  et al. 2013, \pasj, 65, 83

\bibitem[{{Nagase}(1989)}] {n1989}
Nagase, F. 1989, \pasj, 41, 1

\bibitem[{{Narayan \& Yi}(1994)}]{ny1994}
Narayan, R. \& Yi, I. 1994, \apj, 428, L13

\bibitem[{{Narayan \& Yi}(1995)}]{ny1995}
Narayan, R. \& Yi, I. 1995, \apj, 452, 710

\bibitem[{{Negueruela  \& Okazaki}(2001)}]{no2001}
Negueruela, I. \& Okazaki, A.T., 2001, \aap, 369, 108

\bibitem[{{Negueruela et al.}(2001)}]{n2001}
Negueruela, I., Okazaki, A. T., Fabregat, J., et al. 2001, \aap, 369, 117


\bibitem[{{Nomoto}(1984)}] {n1984}
Nomoto, K. 1984, \apj, 277, 791

\bibitem[{{Nomoto}(1987)}] {n1987}
Nomoto, K. 1987, \apj, 322, 206

\bibitem[{{Okazaki}(2001)}]{o2001}
Okazaki, A. T. 2001, \pasj, 53, 119

\bibitem[{{Okazaki et al.}(2002)}]{o2002}
Okazaki, A. T., Bate, M. R., Ogilvie, G. I \& Pringle, J. E., 2002,
\mnras, 337, 967

\bibitem[{{Okazaki et al.}(2013)}]{o2013}
Okazaki, A. T., Hayasaki, K., \& Moritani, Y. 2013, \pasj, 65, 41

\bibitem[{{Okazaki \& Negueruela}(2001)}]{on2001}
Okazaki, A. T. \& Negueruela, I. 2001, \aap, 377, 161

\bibitem[{{Pahari \& Pal}(2012)}]{pp2012}
Pahari, M. \& Pal, S. 2012, \mnras, 423, 3352

\bibitem[{{Parmar et al.}(1989)}]{p1989}
Parmar, A. N., White, N. E., Stella, L., Izzo, C., \& Ferri, P.
1989, \apj, 338, 359

\bibitem[{{Pfahl et al.}(2002)}]{p2002}
Pfahl, E., Rappaport, S., Podsiadlowski, P., \& Spruit, H. 2002, \apj, 574, 364

\bibitem[{{Podsiadlowski et al.}(2004)}]{p2004}
Podsiadlowski, P., Langer, N., Poelarends, A. J. T., et al. 2004, \apj, 612, 1044

\bibitem[{{Poelarends et al.}(2008)}]{p2008}
Poelarends, A. J. T., Herwig, F., Langer, N., \& Heger, A. 2008, \apj, 675, 614

\bibitem[{{Porter \& Rivinius}(2003)}]{pr2003}
Porter, J. M. \& Rivinius, T. 2003, \pasp, 115, 1153

\bibitem[{{Pottschmidt et al.}(2012)}]{p2012}
Pottschmidt, K., Suchy, S., Rivers, E., et al. 2012, in SUZAKU 2011:
Exploring the X-ray Universe: Suzaku and Beyond, ed. R. Petre, K. Mitsuda,
and L. Angelini (AIP Conference Proceedings), Volume 1427, p. 60

\bibitem[{{Pradhan et al.}(2013)}]{p2013}
Pradhan, P., Maitra, C., Paul, B., \& Paul, B. C. 2013, \mnras, 436, 945

\bibitem[{{Pringle \& Rees}(1972)}] {pr1972}
Pringle, J. E. \& Rees, M. J. 1972, \aap, 21, 1

\bibitem[{{Raguzova \& Popov}(2005)}] {rp05}
Raguzova, N. V. \& Popov, S. B. 2005, Astron. Astrophys. Transactions, 24, 151

\bibitem[{{Rajoelimanana et al.}(2011)}]{r2011}
Rajoelimanana, A. F., Charles, P. A., \& Udalski, A. 2011, \mnras,
413, 1600

\bibitem[{{Reig}(2011)}]{Reig2011}
Reig, P. 2011, \apss, 332, 1

\bibitem[{{Reig et al.}(2007)}]{r2007}
Reig, P., Larionov, V., Negueruela, I., Arkharov, A. A.,
\& Kudryavtseva, N. A. 2007, \aap, 462, 1081

\bibitem[{{Reig \& Roche}(1999)}]{rr1999}
Reig, P. \& Roche, P. 1999, \mnras, 306, 100

%

\bibitem[{{Siess}(2007)}] {s2007}
Siess, L. 2007, \aap, 476, 893

\bibitem[{{Shakura \& Sunyaev}(1973)}] {ss1973}
Shakura, N. I. \& Sunyaev, R. A. 1973, \aap, 24, 337


\bibitem[{{Shao \& Li}(2014)}] {sl2014}
Shao, Y. \& Li, X.-D. 2014, submitted


\bibitem[{{Sigut et al.}(2009)}] {s2009}
Sigut, T. A. A., McGill, M. A., \& Jones, C. E. 2009, \apj, 699, 1973

\bibitem[{{Stella, White \& Rosner}(1986)}] {swr1986}
Stella, L., White, N. E., \& Rosner, R. 1986, \apj, 308, 669


\bibitem[{{Townsend et al.}(2011)}] {t2011}
Townsend, L. J., Coe, M. J., Corbet, R. H. D., \& Hill, A. B.
2011, \mnras,  416, 1556

\bibitem[{{van den Heuvel}(2004)}] {vdh2004}
van den Heuvel, E. P. J. 2004, in ESA Special Publication, Vol. 552, 5th
INTEGRAL Workshop on the INTEGRAL Universe, ed. V. Schoenfelder,
G. Lichti, \& C. Winkler (Noordwijk: ESA Publications Division), 185

\bibitem[{{van den Heuvel  \& Rappaport}(1987)}] {hr87}
van den Heuvel, E. P. J. \& Rappaport, S. 1987, in Proc. IAU symposium 98,
Physics of Be stars, eds. A. Slettebak \& T.  P. Snow (Cambridge University Press), p. 291

%

\bibitem[{{Waters \& van Kerkwijk}(1989)}] {wv1989}
Waters, L. B. F. M. \& van Kerkwijk, M. H. 1989, \aap, 223, 196

%
\bibitem[{{White et al.}(1976)}]{w1976}
White, N. E., Mason, K. O., Sanford, P. W., \& Murdin, P., 1976, \mnras,
176, 201

\bibitem[{{Wilson et al.}(2008)}]{w2008}
Wilson, C. A., Finger, M. H., \& Camero-Arranz, A. 2008,
\apj, 678, 1263

\bibitem[{{Wood et al.}(1993)}]{w1997}
Wood, K., Bjorkman, K. S., \& Bjorkman, J. E. 1997, \apj, 477, 926


\bibitem[{{Yi et al.}(1997)}]{y1997}
Yi, I., Wheeler, J. C., \& Vishniac, E. T. 1997, \apj, 481, L51

\end{thebibliography}
\end{document}